\documentclass[10pt,flushrt,tighten,preprint]{aastex}

\newcommand{\erg}{\mbox{$\rm\,erg$}}
\newcommand{\kev}{\mbox{$\rm\,keV$}}
\newcommand{\cm}{\mbox{$\rm\,cm$}}
\newcommand{\ev}{\mbox{$\rm\,eV$}}
\newcommand{\gyr}{\mbox{$\rm\,Gyr$}}

\newcommand{\s}{\mbox{$\rm\,s$}}
\newcommand{\mpc}{\mbox{$\rm\,Mpc$}}
\newcommand{\kpc}{\mbox{$\rm\,kpc$}}

\newcommand{\kms}{\mbox{${\rm\,km\,s}^{-1}$}}
\newcommand{\beq}{\begin{equation}}
\newcommand{\eeq}{\end{equation}}

\slugcomment{Astrophysical Journal, in press}

\shorttitle{X-Ray Disk in NGC 1700}
\shortauthors{Statler \& McNamara}

\begin{document}

\title{A 15-Kiloparsec X-Ray Disk in the Elliptical Galaxy NGC 1700}

\author{Thomas S. Statler and Brian R. McNamara}
\affil{Department of Physics and Astronomy, 251B Clippinger Research
Laboratories, Ohio University, Athens, OH 45701, USA}
\email{statler@ohio.edu, mcnamarab@ohio.edu}

\begin{abstract}
We present {\em Chandra\/} observations of the young elliptical galaxy
NGC 1700.  The X-ray isophotes are highly flattened between semimajor axes of
$30\arcsec$ and $80\arcsec$, reaching a maximum ellipticity $\epsilon_X
\approx 0.65$ at $60\arcsec$ ($15\kpc$).  The surface brightness profile
in the spectrally soft, flattened region is shallower than that of
the starlight, indicating that the emission comes from hot gas rather than
stellar sources. The flattening is so extreme that the gas
cannot be in hydrostatic equilibrium in any plausible potential. A likely
alternative is that the gas has significant rotational support.
A simple model, representing isothermal gas distributed about
a particular angular momentum, can reproduce the X-ray
morphology while staying consistent with stellar kinematics. The specific
angular momentum of the gas matches that of the stars in the most isophotally
distorted outer part of the galaxy, and its cooling time matches the time since
the last major merger. We infer that the gas was acquired in
that merger, which involved a pre-existing elliptical galaxy with a hot
ISM. The hot gas carried the angular momentum of the encounter, and has
since gradually settled into a rotationally flattened, cooling disk.

\end{abstract}

\keywords{galaxies: cooling flows---galaxies: elliptical and lenticular,
cD---galaxies: ISM---X-rays: galaxies---X-rays: ISM}

\section{Introduction\label{s.intro}}

Much of what we know about the mass distributions in the outer parts of
giant elliptical galaxies comes from X-ray observations of their hot
interstellar media (ISM). Since the initial studies of M87
\citep{Mat78,FLG80}, mass profiles of several dozen systems have been
derived from X-ray data \citep[and references therein, for example]{LW99},
strengthening the case for the ubiquity of dark halos. All
these results rely on the assumption that the gas is
in hydrostatic equilibrium, an assumption that is difficult to
validate. If hydrostatic equilibrium holds, the X-ray surface
brightness must trace the projected potential. Conversely, if
there is no plausible potential consistent with the
morphology, then the gas cannot be in equilibrium. But the
distinction between an implausible potential and one that is merely strange
is not clear-cut. The quintessential example is NGC 720, whose X-ray
isophotes, while rounder than the optical isophotes, are flatter than
the projected stellar potential. \citet{BC94} were able to account for
this difference by postulating a dark halo that was alarmingly flat,
though not altogether implausible. Later studies making use of {\em ASCA\/}
\citep{BC98} and {\em Chandra\/} \citep{Buo02} data have shown that
removing the contribution from stellar sources makes the remaining
diffuse emission rounder, relaxing the constraints
somewhat, but still implying a substantial halo flattening.

In the {\em Chandra\/} era of high resolution, luminous ellipticals
continue to appear round in X-rays, even at small radii. This
comes as some surprise in light of arguments by
Mathews and collaborators that at least some of the gas should be rapidly
rotating, and thus not in true hydrostatic equilibrium. \citet{KM95}
pointed out that, since stellar evolution is a primary
source of gas, the gas should retain the specific angular momentum of
the stars, which is typically only a factor of a few below that required
for rotational support. Cooling gas would therefore need to sink
by only a modest factor in radius before becoming
rotationally supported and forming a disk. These ideas were subsequently
fleshed out in detailed simulations by
\citet[collectively BM]{BM96, BM97}. However, a careful examination of archival
{\em ROSAT\/} data by \citet{HB00} showed no indication of central X-ray
disks, and in fact no significant difference between the distribution of
X-ray and optical ellipticities, leading \citet{BM00} to
propose a combination of heat conduction and mass dropout to explain the
absence of the predicted structures.

In this paper we report on {\em Chandra\/} observations of the
young elliptical NGC 1700. We find that this galaxy's X-ray isophotes are
not merely flatter than the stellar potential, but flatter than the
starlight---so flat, in fact, that the gas cannot be in hydrostatic
equilibrium in any plausible potential. We describe the observations and
reductions in \S\ 2 below and the basic results in \S\ 3. We then
argue in \S\ 4 that the gas is settling into a rotationally flattened
cooling disk, similar to those predicted by BM, but much larger in scale.
The high angular momentum and the long cooling time suggest that the gas was
accreted in the last major merger. These points are reiterated, and
their implications briefly discussed, in \S\ 5.

\section{Observations and Data Reduction\label{s.obs}}

NGC 1700 is a relatively isolated elliptical in a loose group
with the nearest companion, the spiral NGC 1699, at a projected distance of
$100\kpc$. An age of $3 \pm 1 \gyr$ \citep[hereafter B00]{Brown00} is
inferred from photometric fine structure \citep{SS92},
stellar dynamics \citep[hereafter SSC]{SSC96}, globular cluster colors
\citep[B00]{Whi97}, and absorption line indices (B00). The galaxy
is photometrically
ordinary inside 2 effective radii ($r_e = 14\arcsec=3.5\kpc$ at $51.4\mpc$)
but becomes very boxy farther out, the parallelogram-shaped outer
isophotes accompanied by faint shells. Detailed stellar kinematics
(SSC) show that the galaxy rotates nearly about its apparent minor
axis, with a mean velocity that rises outward into the boxy region.
Dynamical models imply a weakly triaxial oblate potential that is
dark-matter dominated outside $r_e$ \citep[hereafter SDS]{SDS99}.

NGC 1700 was observed on 2000 November 3--4 with the ACIS instrument on
{\em Chandra.} The galaxy was imaged on the
S3 detector, offset from the nominal aimpoint by 1 arcminute in order to
center it on Node 1. The spacecraft roll angle of $59\arcdeg$ placed
NGC 1699 on S2; however, no emission was detected from NGC 1699,
and no other bright or extended sources were seen by any of the active
detectors. We confine this discussion to the data from S3, for
which the total corrected exposure time was 42,798 s.

The data were reduced and calibrated using CIAO 2.1.3 and CALDB 2.7,
as follows: The level 1 event file was
first reprocessed with the CALDB 2.7 gain map and PHA randomization turned on.
The level 2 event file was rebuilt and the `acis\_detect\_afterglow' correction
removed, then filtered to retain S3 events between $0.3$ and $10.0\kev$.
The resulting lightcurve showed no strong background flares, but
substantial weak flaring and a gradual increase in the background level.
We chose not to use `lc\_clean' and `make\_acisbg' to determine
the background because these procedures would have discarded 60--70\% of
the data. Since the galaxy is small, we opted to keep
the full exposure and derive backgrounds from source-free regions on
the same node of S3. Thus only the GTI filters from the pipeline ``flt1''
file were applied to the data and used to calculate the aspect histogram.

Flux calibrated images were derived by computing instrument and exposure
maps at single energies spaced by $100\ev$, then extracting
$10\ev$-wide energy slices from the event file, calibrating each slice
with the nearest maps, and summing the results. We produced
flux and counts images in three energy bands, soft ($0.3$--$0.8\kev$),
hard ($0.8$--$2.7\kev$), and full ($0.3$--$2.7\kev$).
Visual inspection of images
with different energy cuts showed no noticeable emission from the galaxy above
$2.7\kev$. The division at $0.8\kev$ was chosen to put approximately
equal numbers of counts in the soft and hard bands.
We also produced a $0.5$--$2.0\kev$ image to compare with {\em ROSAT\/}
and {\em ASCA\/} data on other systems.  Adaptively
smoothed images from the inner 512$\times$512 pixel region were
derived using the `csmooth' task. The smoothing
map was determined by running the task on the counts image, with the
minimum and maximum $S/N$ per smoothing element set to 4 and 5 respectively.
The task was then rerun on the flux image using this map.

Profiles of surface brightness and isophotal ellipticity were extracted
from the counts images by three different methods. In the first, the
full-resolution images were further binned at scales of 2$\times$2,
4$\times$4, 8$\times$8, and 16$\times$16 pixels, and elliptical isophotes
were fitted using the `ellipse' task in STSDAS. The center was held fixed
at the centroid of the 4 brightest pixels in the full band image, and the
ellipticity was a free parameter. The major axis position angle (PA) was
taken to be either freely variable or fixed at $90\arcdeg$, the two choices
giving similar results; we show below the results for fixed PA.
At each binning scale, converged fits were found over a range of semimajor
axis ($a$). The results agreed to within the statistical errors where the
ranges overlapped. The final ellipticity and PA profiles were pieced
together so as to sample the profiles as densely as possible while
ensuring that the data points remain statistically independent. In the second
approach, the `ellipse' task was run on the adaptively smoothed images.
We consider this a dangerous alternative because the smoothing scale can
vary along a fitted isophote, and consecutive points in the profiles are not
independent. In the third approach, we applied an iterative
algorithm that diagonalizes the spatial second-moment tensor of the photon
counts in a thin elliptical ring and manipulates the axis ratio until
the eigenvalues match
those for a constant density ring of the same shape. This algorithm was
applied to the unbinned, unsmoothed, full-resolution image. Its disadvantage
is that errors in the fitted parameters can be determined only by Monte Carlo
simulation. All three approaches gave results that are consistent within the
Poisson noise, and so we have chosen to present the results from the first
approach, where the statistical errors are best understood.

Spectra were extracted from various regions in the galaxy
and from two square background regions near the ends of
Node 1 on S3. Because the galaxy is faint---about 3400 counts
in the full band image---spectra from different subregions are not
distinguishable, and so we discuss only the total spectrum extracted
in a $90\arcsec$ circle, excluding three point sources. Single response
matrix and auxiliary response files were computed at the center of the
galaxy image and used for both galaxy and background spectra.

\section{Results\label{s.results}}
\subsection{Flux and Luminosity\label{s.flux}}

\begin{figure}[t]
\epsscale{.7}
\plotone{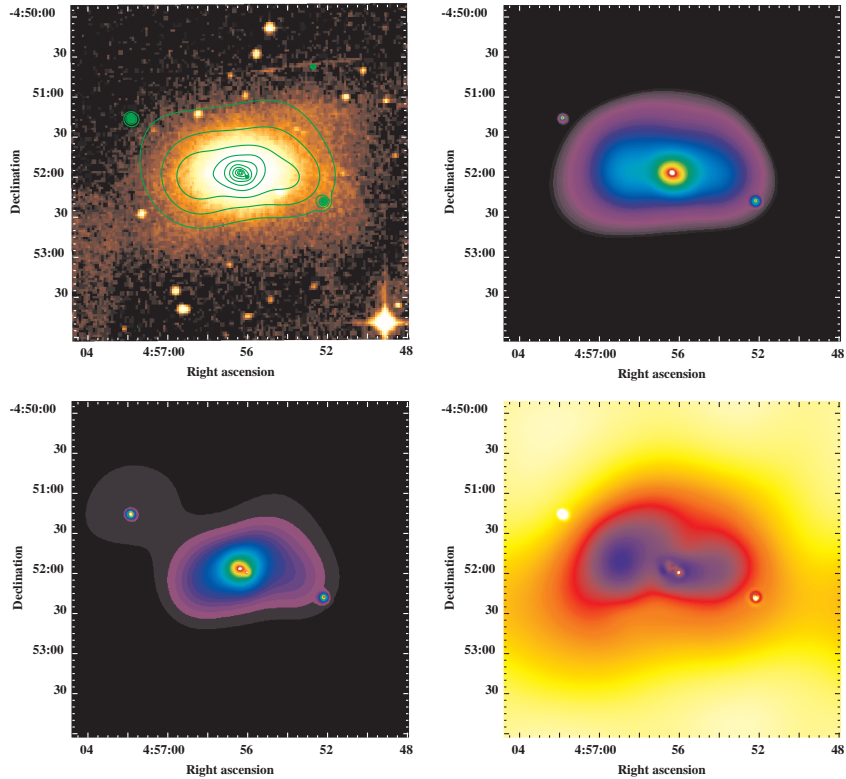}
\caption{\footnotesize
Adaptively smoothed, flux-calibrated X-ray images of NGC 1700:
({\em a; top left\/})
Contours of the full band ($0.3$--$2.7\kev$) image overlaid on the
optical Digital Sky Survey image. In the DSS image, the apparent structure
along the left edge and the streak northwest of center are plate
defects, but the boxiness of the galaxy is real. ({\em b; top right\/})
Soft band image ($0.3$--$0.8\kev$) showing strongly disk-like emission
$20\arcsec$--$50\arcsec$ from the center. ({\em c; bottom left\/}) Hard band
image ($0.8$--$2.7\kev$). ({\em d; bottom right\/}) Hard/soft flux ratio
image; the ratio spans a range from $0.93$ (blue) to roughly $3.2$ in
background-dominated regions (yellow). The central ``squiggle''
is a smoothing artifact.
\label{f.images}}
\end{figure}

The total flux in the full band is $F_{0.3{\rm -}2.7} =
(2.64 \pm 0.10) \times 10^{-13} \erg\cm^{-2}\s^{-1}$ ($1\sigma$ errors).
At a distance of $51.4\mpc$ [$H_0=75\kms\mpc^{-1}$;
\citet{Whi97}], this corresponds to a luminosity $L_{0.3{\rm -}2.7}
=(8.37 \pm 0.32) \times 10^{40} \erg\s^{-1}$. These results include a
correction for Galactic absorption computed from the best-fit spectral
model described in \S\ \ref{s.spec} below.

To compare with existing data,
we use a bandpass of $0.5$--$2.0\kev$, to agree with \citet{BB98}.
Since there is virtually
no flux above $2.0\kev$, this choice also matches the $0.5$--$10.0\kev$
{\em ASCA\/} bandpass \citep{MOM00}. We find
$F_{0.5{\rm -}2.0} = (1.94 \pm 0.06) \times 10^{-13} \erg\cm^{-2}\s^{-1}$,
implying $L_{0.5{\rm -}2.0} = (6.12 \pm 0.19) \times 10^{40} \erg\s^{-1}$.
At this distance, the blue luminosity is $L_B = 6.5 \times
10^{10} L_\sun$. This puts NGC 1700 (with a correction for different
assumed $H_0$) less than $1\sigma$ below the mean $L_{\rm X}$-$L_B$ relation
of \citet{BB98}.

\subsection{Morphology\label{s.morph}}

Figure \ref{f.images}a shows contours of the adaptively smoothed, full-band
image, plotted on the optical image from the Digital Sky Survey. The optical
image shows the transition from the elliptical, slightly disky, inner
region to the parallelogram-shaped outer region, beginning roughly
$30\arcsec$ from the center. Deeper imaging \citep{Brown00} shows the
boxiness reaching a maximum and apparent tidal fine structure appearing
at radii near $100\arcsec$.
The overlaid contours show the striking X-ray morphology
of this system: between an elliptical inner zone and a somewhat boxy outer
zone, the X-ray isophotes are highly flattened---far flatter than the
starlight---over a factor of 2 in radius. The flattening is most prominent
in the soft emission, as can be seen by comparing the
smoothed soft-band and hard-band images in Figure \ref{f.images}b and
\ref{f.images}c. The ratio of the hard to the soft image is shown in Figure
\ref{f.images}d. The ratio image reinforces the basic result that the
softest emission is the flattest emission; the most flattened regions
coincide with the lowest hardness ratios.

\begin{figure}[t]
\vspace{-3pt}
\epsscale{.5}\plotone{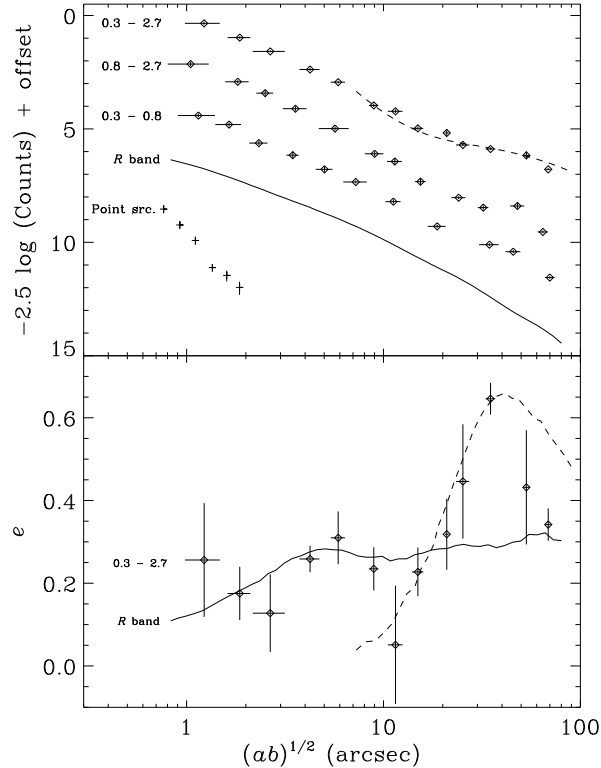}
\caption{\footnotesize
Surface brightness ({\em top\/}) and isophotal ellipticity
({\it bottom\/}) profiles of NGC 1700, plotted in terms of isophotal
mean radius $a(1-\epsilon)^{1/2}$. Points with error bars show
X-ray data in the soft, hard, and full bands as indicated. Horizontal error
bars indicate the spatial bin size (see \S\ \ref{s.obs}). Solid curves
show optical $R$-band profiles for comparison. The profile of a point source
is also shown in the upper plot. Dashed curves indicate fitted profiles
using the model described in \S\ \ref{s.disk}.
\label{f.profiles}}
\end{figure}

The photometric profiles are shown in figure \ref{f.profiles}. The surface
brightnesses in the soft, hard, and full bands are plotted in the upper
panel, along with the $R$ band profile \citep{FIH89}
for comparison. The bottom panel shows
the isophotal ellipticities in the X-ray (full band) and optical. The
profiles show a natural division at a mean radius $a(1-\epsilon)^{1/2}
\approx 20\arcsec$ separating an
inner region, in which the X-ray surface brightness falls off
slightly more steeply than the starlight and the X-ray and optical
ellipticities are nearly equal, from an intermediate-radius region
where the X-ray surface brightness levels out and the ellipticity
increases to $\epsilon_X \approx 0.65$. One can also delineate an outer
region at mean radii beyond $50 \arcsec$, where the surface brightness again
falls and the ellipticity returns approximately to that of the starlight.
In the inner region, the soft, hard, and full profiles can be
fitted by $\beta$ models with $\beta = 0.45$, $0.49$, and $0.47$, respectively.
The difference is statistically significant; the hardness ratio
increases {\em inward\/}, with $d\ln(SB_{0.8{\rm -}2.7}/SB_{0.3{\rm -}0.8})
/d\ln a = -0.23 \pm 0.03$. At the center, there is no indication of an
unresolved nuclear source, as can be seen by comparing the galaxy profiles
with the point source profile in the top of figure \ref{f.profiles}.

The four point sources in Figure \ref{f.images}a may or may not be
associated with NGC 1700. The source nearest the center
falls within $0\farcs2$ of one of the globular cluster candidates
(\#22) identified by \citet{Whi97}, if we assume that the X-ray and optical
centers coincide. However, this source is invisible in the soft band,
indicating that it is probably highly absorbed. Since the upper limit on
NGC 1700's \ion{H}{1} mass is $3.5 \times 10^9 M_\sun$ \citep{Huc94},
we surmise that this source is more likely a background AGN. The remaining
sources, at the distance of NGC 1700, would have luminosities between
$5 \times 10^{38}$ and $5\times 10^{39}\erg\s^{-1}$, comparable with
the brightest sources resolved in NGC 4697 \citep{SIB01}.
On the other hand, the density of sources is similar to
that in the rest of the field, so there is no compelling
reason to conclude that they cannot be background objects.

\subsection{Spectrum\label{s.spec}}

We use a variety of models in XSPEC to fit the total spectrum.
Satisfactory fits ($0.98 < \chi^2/{\rm d.o.f.}
< 1.1$) are obtained with absorbed MEKAL models, while both power-law and
cooling flow models fail. In the absorbed MEKAL fits, the
Galactic \ion{H}{1} column is fixed at
$N_{\rm H} = 4.8 \times 10^{20}\cm^{-2}$ \citep{DickLock}. Because the ACIS-S
calibration is still somewhat uncertain at low energies (M.\ Wise and
C.\ Sarazin, private communications), we apply cutoffs to the spectrum at
either $0.3$ or $0.5\kev$. The free parameters are the gas temperature $T$
and metallicity $Z$; in some fits the redshift is also allowed to vary
to correct for residual energy calibration errors. We find that the systematic
uncertainties are on the order of the statistical errors. The grid of
models results in temperatures in the range $0.44$--$0.49\kev$ and
metallicities in the range $0.47$--$0.55 Z_\sun$, with statistical errors
of $0.02$ and $0.2$ respectively. We conservatively estimate
$T = (0.47 \pm 0.03)\kev$ and $Z = (0.5 \pm 0.3) Z_\sun$.

\section{Discussion}
\subsection{Can the Gas be in Hydrostatic Equilibrium?}

The X-ray emission from elliptical galaxies comes from a mixture of stellar
sources and diffuse gas, in varying proportions. Inside $r_e$,
NGC 1700's X-ray isophotes have the same
ellipticity as the starlight, hinting that stellar sources may
dominate. But we do not see the hard spectral component
characteristic of low-mass X-ray binaries. We therefore cannot
determine whether stellar sources are important in
the inner regions. Farther out, the situation is
different. The X-ray isophotes reach ellipticities $\epsilon_X \approx 0.65$
at a major axis distance of approximately $60\arcsec$ ($15\kpc$). At these
radii, the optical isophotes have $\epsilon_{\rm opt} \approx 0.3$, and the
X-ray surface brightness falls off much less steeply than the starlight.
There is little doubt that this emission is coming from a diffuse ISM.

All studies using X-ray emission to constrain galaxy mass profiles
rely on the assumption that the diffuse ISM is in hydrostatic equilibrium.
Regardless of the equation of state, this condition requires that the
surfaces of constant density, pressure,
temperature, and consequently X-ray emissivity all coincide with
the equipotentials, and therefore that the X-ray surface brightness
traces the shape of the projected potential.\footnote{Because there is only
one plane normal to a given line through a given point, the level
surfaces of two functions must coincide if and only if their gradients are
everywhere parallel (or antiparallel) to each other. The hydrostatic equation,
$\nabla P + \rho \nabla \Phi = 0$, implies that the level surfaces of
$P$ and $\Phi$ coincide. Taking the curl gives $\nabla \rho \times \nabla \Phi
= 0$; thus the $\rho$ and $\Phi$ level surfaces also coincide. $P$ must
therefore be expressible as $P(\rho)$, and any equation of state
$P=P(\rho,T)$ then implies that $T=T(\rho)$. (We
thank the referee for this demonstration.)}

We can test the assumption of hydrostatic equilibrium in NGC 1700
using axisymmetric models, since the galaxy's stellar velocity field implies a
nearly oblate potential (SDS). First, we
estimate the stellar potential using the approach of
\citet[see \S\ 3.1.1 of SDS for details]{BDI}. The stellar potential has
$\epsilon \approx 0.1$ for lines of sight near the equatorial plane, and
we calculate only the edge-on case, since at lower inclinations the gas
would need to be even flatter than it appears. We then ask whether
any plausible dark matter distribution can be added
so as to give the total potential a flattening $\epsilon_\Phi > 0.6$
at mean radii of $35\arcsec$. Intuition argues that this will
require a dark mass at least as large as the stellar mass, highly
flattened and not too centrally concentrated lest its monopole
term be too large.

The largest flattening for a given dark mass would be produced by a 
dark halo that is a disk. A good example is the
\citet{Kuz56} disk, whose surface density is given by
\beq
\Sigma(R) = \Sigma_0 [1 + (R^2 /r_0^2)]^{-3/2}.
\eeq
The constant
$\Sigma_0$ is the central density and $r_0$ is the radius within which
$\Sigma \sim {\rm constant}$. Since the Kuzmin disk has both a very low central
concentration and a steep outer cutoff, one would expect it to be best
able to produce a substantial flattening over a limited range of radii.
We find that the minimum mass able to generate the
correct flattening is $2 \times 10^{12} (M/L_B)_\star
M_\sun$, where stellar mass-to-light ratios $(M/L_B)_\star
\approx 4$ are needed to fit the kinematics at small radii \citep{BBF92}.
However, this case requires $r_0 \approx 90\arcsec$, which
implies a circular velocity curve rising to $\sim 250 (M/L_B)_\star\kms$ at
$R = 50\arcsec$. This is inconsistent with the observed kinematics
(SSC). No Kuzmin disk of any mass can yield $\epsilon_\Phi \ge 0.6$ if
$r_0 < 50\arcsec$, and all Kuzmin disks with $r_0 > 50\arcsec$ massive enough
to flatten the potential produce rotation curves inconsistent with the stellar
kinematics.

To try to preserve the rotation curve, we replace the Kuzmin disk with
\citet{Sch93} logarithmic potentials of various flattenings. These
potentials have a constant circular velocity at all radii, and
are generated by strictly positive mass densities down to quite
large flattenings. We find, however, that no
Schwarzschild density of any flattening has a potential with $\epsilon_\Phi
> 0.35$, simply because they are too centrally concentrated. Other popular
halo models, such as the NFW \citep{NFW} and \citet{Her90}
models, are more centrally concentrated than the Schwarzschild model and
fail in the same way. Finally, a linear combination of Schwarzschild and
Kuzmin models suffers from the same problem as the pure Kuzmin disks, since
the disk potential must be dominant to provide the bulk of the flattening.

We conclude that there is no plausible potential in which the gas can be
in hydrostatic equilibrium and be as flattened as the X-ray isophotes
indicate. Either the gas is well out of equilibrium, or the equilibrium is
hydrodynamical rather than hydrostatic, with angular momentum---i.e.,
bulk rotation of the gas---playing an important role.

\subsection{An Accreted Disk\label{s.disk}}

At large radii, NGC 1700 shows clear optical signatures of a past merger,
including shells and prominent boxiness
\citep{FIH89,Whi97,Brown00}. In deep exposures
the galaxy becomes parallelogram-shaped.
\citet{FIH89} conjectured that the long diagonal
of the parallelogram marks an inclined outer disk or ring, while
\citet{Brown00} interpreted the brightness enhancement along this
direction as a pair of symmetric tidal tails. SSC 
argued that the entire parallelogram could be produced by a broad,
differentially precessing ring seen in projection. By requiring that this
structure be prominent at large radii but phase-mixed beyond visibility at
smaller radii, SSC estimated a dynamical age of $2$--$4 h^{-1}\gyr$,
consistent with other techniques \citep{Brown00}. 

Since both absorption line indices and globular cluster colors also
indicate few-Gyr ages, NGC 1700's most recent merger probably involved
a substantial amount of cold gas and star formation.
\citet{Brown00} suggest that the merger
involved two spirals, but there is no reason to exclude a merger of
a spiral and a pre-existing elliptical with a hot ISM. Such a
merger would dump a load of
already-hot gas into the common potential well, at an angular
momentum characteristic of the galaxy-galaxy encounter. At sufficiently
low densities, this gas would not be channeled to the center as
the cold gas would, but would instead settle into
a rotationally flattened cooling disk.
We propose that the flattened X-ray structure in NGC 1700 is
exactly such a disk, acquired in the same merger that created the 
optical boxiness and tidal features.

We can demonstrate the plausibility of this scenario with a simple model.
First, imagine a population of collisionless particles dumped into
the galactic potential at a fixed kinetic temperature, distributed about
a finite angular momentum $L_0$. After phase mixing, this population
would take on a phase space distribution function (DF) given by
\beq\label{e.df}
f \propto e^{-\beta E} e^{-\alpha(L_z-L_0)^2},
\eeq
where $E$ and $L_z$ are the energy and $z$ component of angular momentum
(both per unit mass), and $\alpha$ and $\beta$ are constants. The first
factor gives an isothermal distribution in energy, while the
second applies a Gaussian bias around $L_0$.
We adopt an axisymmetric model potential,
\beq\label{e.potential}
\Phi(R,z) = {V_c^2 \over 2} \ln \left[1 + {R^2 + (z/q)^2
	\over R_c^2}\right],
\eeq
where $(R,\phi,z)$ are cylindrical coordinates, $V_c$ is the asymptotic
circular velocity, and $R_c$ and $q$ are the core radius and flattening.
This potential gives a flat circular velocity curve for $R\gg R_c$. The
actual rotation velocity of the material follows from the DF:
\beq\label{e.rotcurve}
\langle v_\phi \rangle = {L_0 R \over R^2 + R_0^2},
\eeq
where $R_0 \equiv (\beta/2\alpha)^{1/2}$. We could calculate
the density by integrating over velocities, but there is no need because
the collisionless model is unrealistic. Its velocity distribution becomes
strongly anisotropic for $R\gtrsim R_0$, while in reality collisions should
isotropize the pressure in, at most, a dynamical time.

In the simplest rotating equilibrium, the gas pressure $P$ and density
$\rho$ should obey the hydrostatic equations with an extra term for the
bulk velocity:
\beq\label{e.hydro}
{\partial P \over \partial R} + \rho \left({\partial \Phi \over \partial R}
-{\langle v_\phi \rangle^2 \over R} \right) = 0 , \qquad
{\partial P \over \partial z} + \rho {\partial \Phi \over \partial z} = 0,
\eeq
Here we are again assuming
axisymmetry, and also neglecting shear viscosity. Notice that the hydrostatic
equations do not tell us how to choose the mean velocity
$\langle v_\phi \rangle$. But if we adopt the rotation curve from the
collisionless model [equation (\ref{e.rotcurve})] with the potential of
equation (\ref{e.potential}), and assume an isothermal
equation of state, then equations (\ref{e.hydro}) can be integrated
analytically to obtain the density. Letting the volume emissivity
$\nu$ then be proportional to $\rho^2$, we find
\beq\label{e.emissivity}
\nu = \nu_0 \exp \left({A \over 1 + R_0^2/R^2} \right)
	\left(1 + {R^2 + z^2/q^2 \over R_c^2} \right)^{-B}.
\eeq
In equation (\ref{e.emissivity}), $\nu_0$ is the central emissivity,
$R_0$ is the radius of the peak of the rotation curve,
$A \equiv (\mu m_{\rm H} / kT) (L_0/R_0)^2$ measures the angular
momentum bias, and
$B \equiv (\mu m_{\rm H} / kT) V_c^2$ (with $\mu$ and $m_{\rm H}$ the mean
molecular weight and hydrogen mass) describes the depth of the potential
relative to the temperature.

We project the emissivity of equation (\ref{e.emissivity}) for a line of
sight in the equatorial plane, and fit elliptical isophotes.
The dashed curves in Figure \ref{f.profiles} show the isophotal profiles
in the relevant range of radii for a model with $A=8.0$, $B=2.3$, and
$R_0=50\arcsec$. The potential is spherical ($q=1$), with $R_c$
set small enough that it does not influence the results. The
model does a credible job of reproducing both profiles, over nearly a
decade in mean radius, despite its extreme simplicity. As a check, we
can use the value of $B$ and the gas temperature to predict the circular
velocity. Assuming $\mu=0.6$, we obtain $V_c = 420 \kms$. Treating the
dark halo as a singular isothermal sphere, this implies a velocity
dispersion $\sigma_{\rm halo}=2^{-1/2}V_c = 290\kms$. The stellar component,
having formed dissipationally, is dynamically colder than the halo and should
have a lower dispersion. If we model both stars and halo as
constant-dispersion systems with power-law density profiles having logarithmic
slopes of $-3.5$ and $-2$ respectively, we predict $\sigma_{\rm stars} /
\sigma_{\rm halo}= 2/7^{1/2}$ \citep{Car94}. This implies $\sigma_{\rm
stars} = 220 \kms$, which compares favorably with the measured
peak dispersion (outside the cold counterrotating core) of $230\kms$ (SSC).

At the radius of maximum flattening ($56\arcsec$),
the gas in the model has a mean 
rotation velocity of $270\kms$, about two-thirds of the circular velocity.
This gives it a specific angular momentum twice as large as that
of the stars at the same radius.
The stellar rotation velocity $V_\ast \approx 150\kms$ at the outermost
measured point ($65\arcsec$). If we assume that $V_\ast$ remains
approximately constant to larger radii, then
the stars carrying the same specific $z$ angular momentum as the gas 
would be found at $R \approx 90\arcsec$, coinciding with the
maximum distortions of the optical isophotes.
We would naturally expect such a concordance if the stars making up
the stellar ring and the X-ray gas were accreted together.

Finally, in order to settle into a disk, the gas must have had time to cool.
At $T\approx 0.5\kev$ and $Z\approx Z_\sun/2$, the cooling function
$\Lambda \approx 6 \times 10^{-15} \kev\cm^3\s^{-1}$ \citep{BH89}.
Normalizing to the observed surface
brightness at $56\arcsec$, we find the electron density in the model
at $R=14\kpc$ to be $3 \times 10^{-3} \cm^{-3}$ in the midplane of the disk.
The cooling time is therefore $t_{\rm cool} \approx 3 \gyr$, which
is equal to the dynamical age of the merger debris. Thus, enough time has
elapsed since the merger for the gas to have partially cooled and settled,
but not so much that it would have sunk completely to the bottom of the
potential well.

\section{Conclusions}

{\em Chandra\/} observations of NGC 1700 show extended X-ray emission that
is dominated by a hot ISM. The gas is distributed in a flattened
structure that is inconsistent with hydrostatic
equilibrium in any plausible potential. A simple model in which gas at a
single temperature, centered around a particular angular momentum, is
dispersed in a logarithmic potential can account for the gross
X-ray morphology and be consistent with stellar kinematics.
The angular momentum of the gas matches that of the stars in the
isophotally distorted outer part of the galaxy, and its cooling time
matches the dynamical age of the last merger. We infer that the
gas was acquired in that merger, carrying the angular momentum of the
encounter, and is settling into a rotationally flattened, cooling disk.

The presence of a rotationally supported disk in this object, and
the growing evidence for shocks and bubbles in the ISM of elliptical
galaxies \citep{Jon01,Fin01} casts 
doubt on the assumption that elliptical galaxies are simple, hydrostatic
systems. Mass profiles derived from X-ray data and hydrostatic models may
therefore be in error, even in those systems that appear symmetric. A modest
degree of rotational support in the gas may significantly affect the shape of
the inferred mass distribution, depending on how the angular momentum is
distributed. In the models of \S\ \ref{s.disk}, rotational
support increases outward interior to $R_0$ and decreases outward beyond
this radius. Applying a hydrostatic model would therefore lead
to a mass profile that is too centrally concentrated. Interestingly, galaxy
cluster mass profiles are found, in some cases \citep{Dav01,Nev01},
to be unusually concentrated compared to the expected NFW \citep{NFW} profile.
Whether this discrepancy could also stem from gas rotation remains to be seen.

\acknowledgments

The authors are grateful to Yousaf Butt, Pat Slane, and the staff of the CXC
for their skillful operation of {\em Chandra\/}, and to Mike Wise, Craig
Sarazin, Duncan Forbes, Joe Shields, and Steven Diehl for helpful suggestions
and comments. Support for this work
was provided by grant GO1-2094X from the CXC.

\end{document}